\documentclass[%
 reprint,
superscriptaddress,
 amsmath,amssymb,
 aps,
]{revtex4-2}

\usepackage{graphicx}
\usepackage{dcolumn}
\usepackage{bm}
\newcommand*\diff{\mathop{}\!\mathrm{d}}
\newcommand{\lb}{\left(}
\newcommand{\rb}{\right)}
\newcommand{\n}{\nonumber}
\newcommand{\bo}{\boldsymbol}
\newcommand{\nv}{\hat{\bm{n}}}
\newcommand{\tv}{\hat{\bm{t}}}
\newcommand{\vc}{\bm{v}_c}
\newcommand{\sinn}[1]{\sin\left( #1 \right)}
\newcommand{\coss}[1]{\cos\left( #1 \right)}

\newcommand{\pv}{\Delta \bm{p}}
\newcommand{\mean}[1]{\left\langle #1\right\rangle}
\newcommand{\br}[1]{\left( #1\right)} 

\usepackage{xcolor}

\newcommand{\az}[1]{\textbf{{\textcolor{blue}{}}}}
\newcommand{\com}[1]{\textbf{{\textcolor{brown}{}}}}
\newcommand{\add}[1]{#1}

\begin{document}

\author{Niklas Grimm}
\affiliation{Fachbereich Physik, Universit\"at Konstanz, 78457 Konstanz, Germany}
\author{Annette Zippelius}
\affiliation{Institut f\"ur Theoretische Physik, Georg-August-Universit\"at G\"ottingen, 37077 G\"ottingen, Germany}
\author{Matthias Fuchs}
\affiliation{Fachbereich Physik, Universit\"at Konstanz, 78457 Konstanz, Germany}

\preprint{APS/123-QED}

\title{Simple fluid with broken time reversal invariance}

\date{\today}

\begin{abstract}
We characterize a system of hard spheres with a simple collision
rule that breaks time reversal symmetry, but conserves energy.
The collisions lead to an a-chiral, isotropic, and homogeneous
stationary state, whose properties are determined in simulations and compared to an approximate theory originally developed for elastic hard spheres.
In the nonequilibrium fluid state, velocities are correlated, a phenomenon known from other nonequilibrium stationary states. The correlations are 
 long-ranged decaying like $1/r^d$ in $d$ dimensions. Such correlations are expected on general grounds far from equilibrium and had previously been observed in driven or non-stationary systems. 
\end{abstract}

\maketitle

\section{\label{sec:level1} Introduction }

Fluids far from equilibrium are receiving increasing interest, driven
mainly by research on active matter~\cite{Marchetti2013,Bechinger2016}
and granular materials~\cite{Behringer2019,Poeschel}. Nonequilibrium
stationary states due to imposed shear or gradients of the temperature
have been studied extensively in the past~\cite{Machta1980,Dorfman1994}. In contrast, active matter
and granular systems are not only recent topics, but also inherently
different: In both systems energy is exchanged locally with the
environment. Active particles extract energy from their environment to
perform certain functions, such as propulsion. In granular media
energy is lost in collisions, possibly to internal degrees of
freedom. Both systems are not symmetric under time reversal and do not
conserve energy.

The consequences of
the broken time reversal invariance on correlations in the nonequilbrium stationary state have been a topic of long standing interest.  Strong long ranged correlations  were observed in systems in external gradients  \cite{Machta1980}. Anisotropic transport processes \cite{Dorfman1994}   and boundary driving  \cite{Eyink1996} were identified as sources of weaker correlations decaying asymptotically as $1/r^d$ in $d$ dimensions. They are absent in equilibrium systems, and generically but not necessarily  expected in nonequilibrium stationary states \cite{Bertini2015}. In granular systems they were predicted in time-dependent (freely cooling) states \cite{Noije1998}, and in active fluids they may be connected to interactions among passive particles \cite{Baek2018}. Yet, the necessary prerequisites for observing long range correlations in nonequilibrium remain unknown. 

Here, we study a simple fluid which respects all conservation laws,
including energy conservation, and yet breaks time-reversal symmetry,
allowing us to disentangle the effects of violation of energy
conservation and breaking of time reversal invariance. Our model is a
coarse grained version of rough spheres~\cite{Chapman}, previously
introduced to model molecules with internal degrees of
freedom~\cite{Candif1965} and granular particles~\cite{Huthmann1997};
in both cases energy can be taken up by internal degrees of freedom
which are not explicitly taken into account. Here we insist on energy
conservation: Roughness just gives rise to a coupling of rotational
and translational degrees of freedom during collisions. Momentum
transfer along normal and tangential direction is balanced, such that
\add{the sum of translational and rotational kinetic} energies is strictly conserved in collisions. \add{Additionally, homogeneity, isotropy and a-chirality are enforced. Perhaps surprisingly, the freedom to model the collision process still opens the possibility to invent scattering laws which run differently forward and backwards in time.} Considering a general planar collision process leaves us with one
free parameter, $\chi$, which controls \add{how strongly the forward-running and backward-running scattering processes differ. It also controls} how rapidly particles move
apart after colliding. For $\chi\to 1$, the particles move in parallel
(sticky limit), for small $\chi$ they move apart much more rapidly than
for smooth spheres. \add{For intermediate $\chi$, a quasi-equilibrium state ensues.} 

We have analysed the stationary state of the model with MD simulations
and analytical theory, based on the Pseudo Liouville operator.  We
show that equipartition does not hold and velocities are correlated as for active matter~\cite{Dombrowski2004,Cavagna2010,Marconi2016,Flenner2016}. 

The correlations are nontrivial, oscillating in space with roughly the
nearest neighbour distance. Depending on the free parameter $\chi$
particles in proximity are positively correlated for large $\chi$ (motion in parallel) and negatively correlated for small $\chi$ (rapid motion apart). {This local correlation survives into the far field, where longitudinal velocity fluctuations decorrelate algebraically. The decay follows  $1/r^d$ asymptotically for $r\to\infty$, with state-dependent amplitude.}

The paper is organized as follows. In Sect.~\ref{sec:II} the scattering law is presented, Sect.~\ref{sec:III} gives details on the simulations and Sect.~\ref{sec:IV} describes the theory with details relegated to an Appendix. Sect.~\ref{sec:V} discusses the violation of the classical equipartition theorem, and Sect.~\ref{sec:VII} presents the stationary velocity correlations, first on distances comparable to the average particle separation, then in the far field.

\section{Model system}
\label{sec:II}

The new collision law couples rotation and translation, while conserving energy, momentum and angular momentum. In general, collisions are not invariant under time reversal. The simulations are run for monodisperse two dimensional disks, while the generalization for 3D multi-disperse systems is straight forward.
Theory is introduced in 3D and then transcribed to 2D.

\subsection{Elastic collision law}

\begin{figure}
    \includegraphics[scale=0.5]{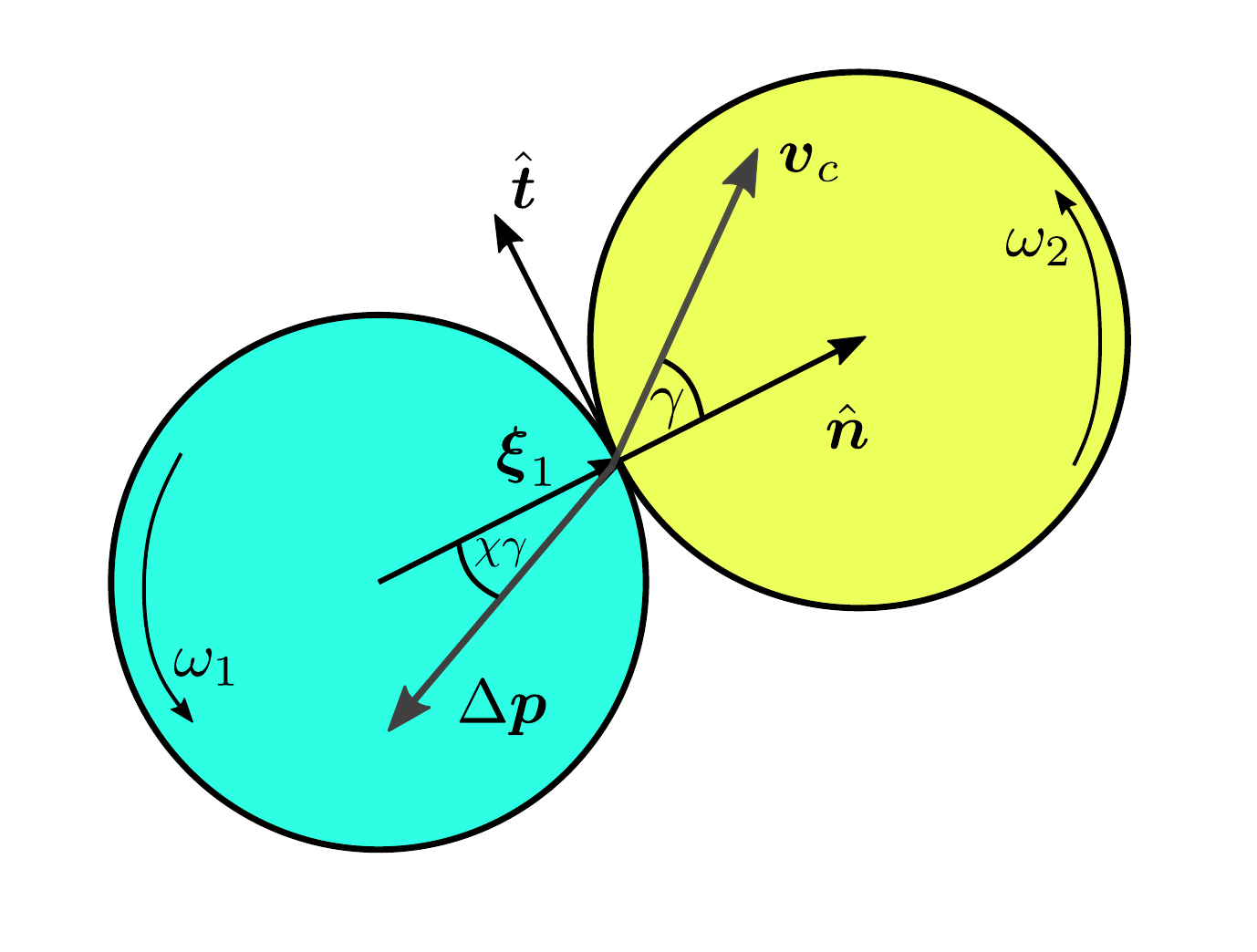}
    \caption{Illustration of the collision rule. The constant parameter $\chi$ governs the direction of the momentum transfer $\pv$. $\chi=0$ recovers smooth spheres with  $\pv$ always along the axis of $\nv$. }
    \label{fig:colEvent}
\end{figure}

The collision of two hard spheres is described by an instantaneous collision event that assigns both particles new velocities $\bm{v}'_i$ \add{and new angular velocities} $\bm{\omega}'_i$, calculated from the spheres \add{(angular)} velocities $\bm{v}_i$  ( $\bm{\omega}_i$ ) prior to the collision and their angle of collision $\gamma$.
The equations of the collision rule for two particles $i=1$ and 2 read
\begin{gather}
    \bm{v}_{1}' = \bm{v}_{1} + \frac{1}{m_1} \pv, \qquad   \bm{v}_2' = \bm{v}_2 - \frac{1}{m_2} \pv , \n \\
    \bm{\omega}_{1}' =  \bm{\omega}_{1} - \frac{R_1}{I_{1}} \left( \pv \times \nv  \right),   \qquad    \bm{\omega}_{2}' =  \bm{\omega}_{2} - \frac{R_2}{I_{2}} \left( \pv \times \nv \right), \label{eq:law1}
\end{gather}
with the unit vector
$\nv=\lb \bm{r}_{2}-\bm{r}_1 \rb/\left| \bm{r}_{2}-\bm{r}_1 \right|$
pointing along the collision axis. $R_j$, $I_j$ and $m_j$ denote the
radius, moment of inertia, and mass of sphere $j$, respectively. An exemplary
collision event is shown in Fig.~\ref{fig:colEvent}, \add{ where also $\bm{\xi}_j$, the vector pointing from the center of sphere $j$ to the contact point, is marked. Also the tangential vector} pointing in the direction of the tangential component of $\bm{v}_c$ in the collision plane is marked; it is defined as
\begin{gather}
    \tv = \frac{(\nv \times \bo{v}_c) \times \nv}{|(\nv \times \bo{v}_c) \times \nv|}.
\end{gather}
The above structure in Eq.~\eqref{eq:law1} already incorporates the conservation of linear momentum and angular momentum $\bm{L} = \bm{r} \times \bm{p} + I \bm{\omega}$,
consisting of orbital momentum and spin. To simplify, the change in the internal angular momentum has been taken to arise solely from the momentum transfer; a possible instantaneous torque is neglected.
\\

The relative velocity at the contact point is given by
\begin{eqnarray}
 \bm{v}_c =&& (\bm{v}_1 + \bm{\omega}_1 \times \bm{\xi}_1 ) -   (\bm{v}_2 + \bm{\omega}_2 \times \bm{\xi}_2 ) \nonumber \\
 =&& \bm{v}_1 - \bm{v}_2 + \left( R_1 \bm{\omega}_1 + R_2 \bm{\omega}_2 \right) \times \nv, \label{eq:vc}  
\end{eqnarray}

and the corresponding quantity after collision reads from Eq.~\eqref{eq:law1}

\begin{gather}
  \bm{v}_c'= \bm{v}_c+\frac{1}{\mu} \pv + \frac{q}{\mu}
  (\Delta \bm{p}- \pv_n)
    \label{eq:vcp}
\end{gather}
with
$1/\mu = 1/m_1 + 1/m_2$ and $q = \mu (R_1^2/I_1 + R_2^2/I_2)$.  The
component of $\pv$ along the normal direction is denoted as
$\pv_n=(\Delta \bm{p}\cdot\nv)\nv$.\\

We require the kinetic energy,
\begin{equation}
E_{kin}=\sum_{i=1}^N\big(\frac{m_i}{2}
\bm{v}_i^2+\frac{I_i}{2}\bm{\omega}_i^2\big)
\end{equation}
to be conserved in collisions.
The difference of the energies before and after the collision can be calculated to
\begin{eqnarray}
2 \Delta E_{kin} =&& \Delta \bm{p} \cdot  [  \bm{v}'_1 + \bm{v}_1 - \bm{v}'_2 - \bm{v}_2  \nonumber \\ 
+&&   \lb R_1 \bm{\omega}_1 + R_1\bm{\omega}_1' + R_2\bm{\omega}_2 + R_2\bm{\omega}_2' \rb \times \nv]  \n \\
=&& \Delta \bo{p} \cdot \left[ \bm{v}_c' +\bm{v}_c \right] . 
\end{eqnarray}
For elastic collisions we demand $\Delta E_{kin}=0$, implying for the momentum transfer \add{(with Eq.~\ref{eq:vcp})} 
\begin{eqnarray}
2 \mu \Delta \bm{p} \cdot \bo{v}_c + \Delta \bm{p}^2 + q |\Delta \bm{p} \times \hat{\bm{n}}|^2 = 0. \label{eq:en0}
\end{eqnarray}

It is important to note that the conservation laws do not determine
the momentum transfer uniquely.  Two well studied cases are, $(i)$ smooth spheres, where collisions do not affect the rotations, $\Delta \bm{p}=\Delta \bm{p}_n$, and $(ii)$ rough spheres, where the contact velocity gets reflected, $ \bm{v}_c'=- \bm{v}_c$ \cite{Bryan1894}. In general,  one scalar equation,
$\Delta E_{kin}=0$, provides one relation between the 3 components of
the momentum transfer in 3 dimensions, leaving us with two degrees of
freedom, as  noted by~\cite{Crawford1989,Meanwell2017}. In the following, we assume a planar scattering geometry, so that the momentum transfer can be
decomposed into a normal and a tangential component (see Fig.~\ref{fig:colEvent})

\begin{gather}\label{eq:deltap1}
        \Delta \bm{p} = - \left| \Delta \bm{p}  \right| \left( \cos\left( \chi \gamma  \right)  \hat{\bm{n}} +\sin\left( \chi \gamma  \right)  \hat{\bm{t}} \right),
      \end{gather}
      leaving us with one free parameter $\chi$~\add{\footnote{The scattering plane is spanned by the relative velocity before the collision  $\bm{v}_c$ and the normal vector $\nv$ pointing between  the sphere centers. }}.
      Substituting the expression of Eq.~\eqref{eq:deltap1} into Eq.~\eqref{eq:en0}, we can solve the equation for $|\pv|$ and obtain

\begin{gather}\label{eq:deltap2}
\left| \Delta \bm{p} \right| = \frac{ 2 \mu   \cos \left( \gamma \chi - \gamma  \right) }{1 + q \sin^2\left( \chi \gamma  \right) } \left| \bm{v}_c \right| .
\end{gather}

The above representation of the momentum transfer, reveals that the
free parameter $\chi$ controls the coupling of the translational and
rotational degrees of freedom. We may thus consider our model as a coarse grained version of rough spheres. In contrast to most other work on rough spheres, our model conserves energy by properly adjusting normal and tangential momentum transfer. This is seen in the limit of small $\chi$, where
Eqs.~(\ref{eq:deltap1},\ref{eq:deltap2}) reduce to
\begin{align}
  -\Delta \bm{p} = &\nv m|v_c|\cos{(\gamma)} \nonumber  \\ &+  \chi\gamma m|v_c|(\nv \sin{(\gamma)}  +  \tv \cos{(\gamma)} ) + {\cal O}(\chi^2) \label{eq:deltap3}.
\end{align}
The first term on the right hand side is just the reversal of the
normal component, well known from smooth elastic spheres. The second
term is controlled by $\chi$ and represents normal as well as
tangential momentum transfer which, however, is not independent as for
inelastic rough spheres. Instead, the two components of momentum transfer are connected by energy conservation. In the opposite limit $\chi=1$, the momentum transfer $\Delta \bm{p}$ is antiparallel to the precollision relative velocity  $\bm{v}_c$, as can be seen from Fig.\ref{fig:colEvent}. 

In the following, we specialise to a monodisperse system \add{with diameter $d=2R$}, so that
the discussion can be simplified with $2 \mu  \equiv m$ and $q \equiv \alpha=m R^2/I$, implying $\alpha=1$ for rings and $\alpha=2$ for discs~\footnote{Homogeneous disks with $\alpha=2$ are called 'discs' throughout the text to discern them from the general case.}.
\add{In order to simplify the system further,} we consider a two dimensional fluid, such that translation is only allowed in $x$- and $y$-directions, while $\omega$ is the angular velocity component along the $z$-axis. 
Units of energy are chosen, such that $k_B=1$ and the time is measured in units of $t_0 = \sqrt{\frac{m d^2}{T_0}}$. $T_0$ denotes the Temperature of the Maxwell-Boltzmann distribution the system is initialized with. For smooth disks (SD) the only thermodynamic control parameter is the packing fraction $\phi=(N/V)\,\pi R^2$.

\subsection{Symmetries} 

\begin{figure}
    \centering
    \includegraphics[scale=1.0]{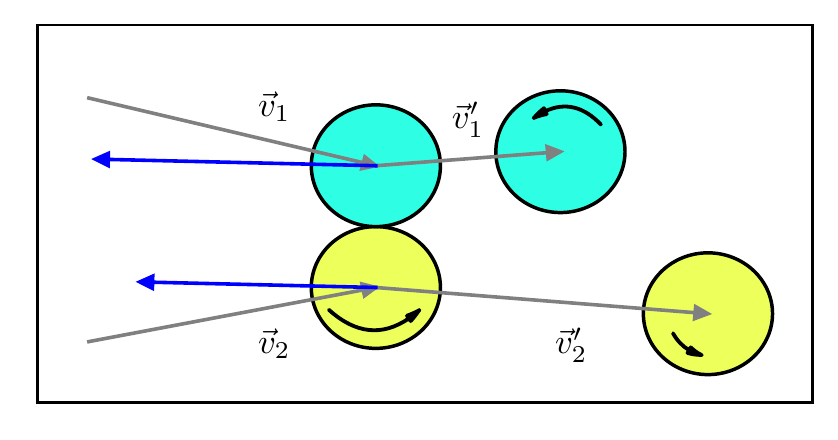}
    \caption{A collision event of two disks  and the time-reversed process are shown; rings, $\alpha=1$ and $\chi=1.0$ are chosen. Gray and curved black arrows denote the velocities and angular velocities, respectively, of the collision running forward in time.  When time is reversed, the particles move in parallel after the reversed collision; the final velocities of the reversed process are marked by blue arrows. We refer to this parallel motion as sticky behavior. Since initial gray and reversed blue arrows are not anti-parallel, this scattering process breaks time reversal symmetry.}
    \label{fig:traj}
\end{figure}

The scattering law is Galileian invariant since only differences of the velocities enter. 
It is also invariant under arbitrary rotations including mirroring and inversions, denoted by an orthogonal matrix $\mathcal{R}$\add{, as can be shown as follows:}
The angle $\gamma$ can be calculated from the ratio between the normal component $v_n = \bm{v}_c \cdot \nv$ and $|\bm{v}_c|$
\begin{gather}
    \gamma =\arccos \left( \frac{v_n }{|\bm{v}_c|} \right), \label{eq:gammaDef}
\end{gather}
where the scalar product is invariant by orthogonal transformations and so is $\gamma$. Since $\nv$
and $\tv$ are vectors and transform as such, the momentum transfer transforms as \add{vector under rotation and reflection}  $\pv \rightarrow \mathcal{R} \cdot \pv$. 

The scattering law is not symmetric under time reversal. \add{Let }  two particles $j=1,2$ collide with $\bm{v}_j$ and $\bm{\omega}_j$ \add{so that} after the collision \add{they move away from each other with} $\bm{v}'_j$ and $\bm{\omega}'_j$, as shown in Fig.~\ref{fig:traj}. If the time is reversed, the momenta are reversed, i.e. $\bm{v}'_j \rightarrow \bm{v}^r_j = -\bm{v}_j'$ and $\bm{\omega}'_j \rightarrow \bm{\omega}^r_j= -\bm{\omega}_j'$, \add{which leads to a precollisonal velocity $\bm{v}_c^r$ of the time-reversed scattering event.} This reversed $\bm{v}_c^r$ is not antilinear to $\bm{v}_c$ and therefore the momentum transfers differ, $\pv^r \neq - \pv$.
Most easily, this can be shown by calculating the scattering angles $\gamma \neq \gamma^r$ that are given by Eq.\,\eqref{eq:gammaDef}. It is enough to only evaluate the argument with Eq.~\eqref{eq:vcp}, which yields 
\begin{gather}
\frac{\bm{v}_c^r\cdot \nv }{|\bm{v}_c^r|} = \frac{-v_n - \frac{2}{m} \pv \cdot \nv }{|-\bm{v}_c - \frac{2}{m} \pv +\frac{2R^2}{I}\br{ \pv \times \nv }|}.
\end{gather}
This expression is equal to $v_n /|\bm{v}_c|$ of the forward collision only if the particles collide centrally and without angular velocities, i.e. $\pv=m (\bm{v}_2 - \bm{v}_1)= \pv \cdot \nv$ in that special case.

All discussed symmetry properties of the scattering law hold in 2D and 3D.
An example of a scattering event of two rings and its time-reversed complement is shown in Fig.~\ref{fig:traj} for $\chi=1$. Here, parameters are chosen such that the time-reversed scattering process leads to a vanishing normal component of the postcollision relative velocity which may be called a 'perfectly sticky collison'.

\subsection{Sticky collisions}
\label{sec:sticky}

\begin{figure}
    \centering
    \includegraphics{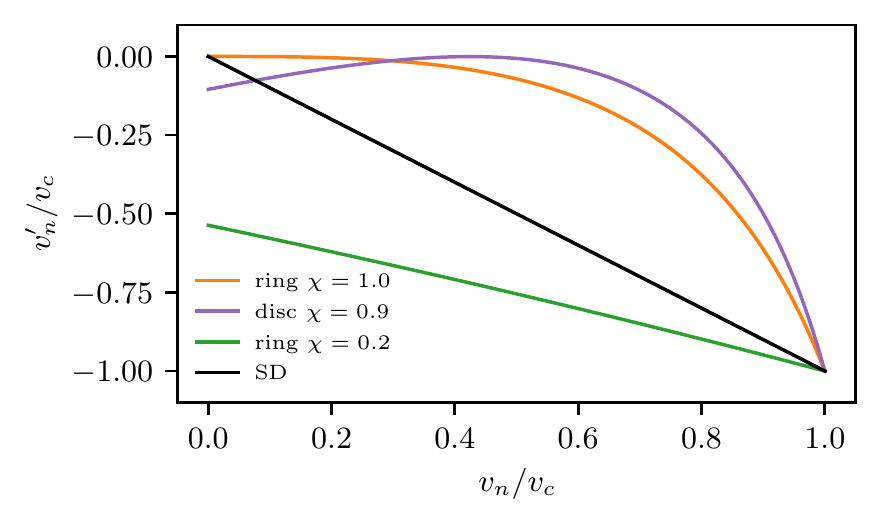}
    \caption{Ratio of $v_n'/v_c$ after collision depending on $v_n/v_c$ before collision for a fixed $v_c>0$. 
      The curve for $\chi=0$ is familiar from SD, where $v_n$ just gets reflected. For $\chi$ close to 1, $v_n'$ is close to 0 even 
      for $v_n/v_c \approx 0.5$, implying (almost) sticky collisions.}
    \label{fig:vn}
\end{figure}

In the case of $\chi$ being strictly zero, $\chi=0$, the scattering is the elastic collision law
of smooth disks (SD)
without rotation-translation coupling. For all
$\chi > 0$, it is helpful to consider the translational motion of two
particles after they have collided. The normal component of the
relative velocity at contact before collison reads
$v_n = \nv \cdot \vc = \nv \cdot(\bo{v}_1 - \bo{v}_2)$ and after
collision from Eq.~\eqref{eq:vcp} follows
\begin{equation}
    v_n' = v_n+\frac{2}{m} \Delta \bo{p} \cdot \hat{\bo{n}}  = v_n -  \frac{2 v_n \coss{\chi \gamma}\coss{\gamma \chi - \gamma } }{ \coss{\gamma} \lb 1 + \alpha \sin^2\left( \chi  \gamma \right) \rb  }  \leq 0, \label{eq:sticky}
\end{equation}
where the last inequality enforces that the particles do not overlap.
For the particles to collide $v_n > 0$ has to hold before the
collision. The case that $v_n' = 0$ is possible for some $\gamma$ and
this implies, that the particles move in parallel after the
collision. Eq.\,\eqref{eq:sticky} can be divided by some arbitrary
$v_c>0$ to generate Fig. \ref{fig:vn}. There the black curve for
SD shows the linear behavior of simple reflection. The curves
for $\chi=1$ and $\chi=0.9$ give small values for $v_n'/v_c$ over a
wide range, in which the colliding particles may stay in proximity after
the collision.
The purple curve ($\chi=0.9$) even shows non-monotonic
behavior, changing slope where particles have a large $v_n/v_c\approx 0.5$ before the
collision and possibly stay close together after it. On the other hand, small $\chi$ give rise to values of  $|v_n'|>|v_n|$, implying that the particles move apart with larger relative velocity than before collision.

To guarantee hard sphere interactions the inequality of  Eq.~\eqref{eq:sticky} has to hold for all $\gamma$, which is not the case for arbitrary system parameters $\chi$ and $\alpha$. The range of $\chi$-values, from which $\chi$ can be chosen, has an upper limit, depending on $\alpha$. For the smallest reasonable $\alpha=1$, i.e.~rings where all mass is at the particles surface, $\chi \leq 1$ has to hold.  With increasing $\alpha$ the maximal possible $\chi$ decreases.\\

\section{Simulation system}
\label{sec:III}
We perform event driven simulations of disks moving in two dimensions \cite{edbdScala}.  The simulation starts with drawing random velocities from a Maxwell-Boltzmann distribution at temperature $T_0$ and then lets the system equilibrate into its stationary state. 
For two translational and one rotational degrees of freedom the energy in the system is $E=\frac{3}{2} N T_0$.
Since energy is conserved, we avoid using a thermostat. 

Cases with perfectly sticky collisions like in Fig.~\ref{fig:traj} are a problem to the algorithm, because the particles would collide continuously \add{during their parallel motion}. Therefore\add{, for simulations at finite $\phi$} we choose parameter pairs of $\chi$ and $\alpha$ that do not allow perfect sticky collisions. For example the orange curve in Fig. \ref{fig:vn} would produce continuous collisions in cases for small $v_n$. We use $\chi=0.99$ instead of $\chi=1$ for the simulations to have more stable equations. More simulation details can be found in the appendix Sec.\,\ref{sec:AppSim}.

\section{Liouville Theory}
\label{sec:IV}

{An approach to describe the macroscopic state variables in the stationary nonequilibrium state can build on kinetic theory \cite{Dorfman2021} and the theory of fluids \cite{Hansen}.}
Following previous work on hard sphere dynamics, the scattering law can be implemented in a Liouville dynamics~\cite{Ernst1969,Resiboisa1975,Resiboisb1975,Leutheusser1982,Huthmann1997,Aspelmeier2001}. 
The time-dependence of a general phase space variable is given by
\begin{gather}
    A\left( \left\{ \bm{v}_i, \bm{r}_i, \bm{\omega}_i   \right\}, t \right) = e^{i \mathcal{L}_+ t}  A\left( \left\{ \bm{v}_i, \bm{r}_i, \bm{\omega}_i   \right\}, 0 \right).
\end{gather}
Here $\mathcal{L}_+=\mathcal{L}_0+\mathcal{L}_+'$ is the pseudo Liouville operator to go forward in time. It consists of the free streaming part, 
where we require the translation term $i\mathcal{L}_0 = \sum_j\bm{v}_j \cdot (\partial / \partial \bm{r}_j)$ only, and the collision part $\mathcal{L}_+'$, which reads: \\
\begin{gather}\label{coll_op}
    i \mathcal{L}_+' = \frac{1}{2} \sum_{j \neq k} \mathcal{T}_{jk} \\
    \mathcal{T}_{jk}= - (\bm{v}_{jk} \cdot \hat{\bm{r}}_{jk})\Theta(- \bm{v}_{jk}\cdot \hat{\bm{r}}_{jk} ) \delta (r_{jk} - R_j - R_k)(b_{jk}^+-1)\nonumber.
\end{gather}
Here and in the following, $\bm{r}_{ij}$ denotes $ \bm{r}_{ij}= \bo{r}_i - \bo{r}_j$ and $\bm{v}_{ij} = \bo{v}_i - \bo{v}_j$.
The exchange operator 
\begin{gather}
    b_{kl}^+ a(\bm{r}_i, \bm{v}_i, \bm{\omega}_i ) =  a(\bm{r}_i, \bm{v}_i', \bm{\omega}_i' )  \qquad  \text{for} ~ i \in \{k,l\},
\end{gather}
replaces precollisional velocities by postcollisional ones for colliding particles while leaving the rest unchanged. \add{Here, $a$ represents a general function depending on positions and velocities}.

It is not possible to compute the $N$-particle distribution function,
$\rho$, from first principles. \add{Based on observing steady state properties in the simulations, we conjecture  ergodicity, viz. that the pdf approaches a stationary $\rho$ which satisfies $ i \overline{\mathcal{L}_+}\,\rho=0$; see Ref.~\cite{Aspelmeier2001} for the construction of the pdf-pseudo Liouvillean  $ i \overline{\mathcal{L}_+}$. 
It follows that $\rho=\rho(\Gamma;E,N,V,\chi,\alpha)$. Determining it further is difficult because particle velocities and positions become correlated in the scattering events; 
this is shown e.g.~in Fig.~\ref{fig:vvr} below.}

In order to compute temperature and
pressure, we \add{can exploit that the averages contain the collision Liouville operator from Eq.~\eqref{coll_op}. Thus, the correlation between velocities and particle separations is only required at contact, viz.~for $r_{ij}=d$.  Additionally, we have checked (not shown)  that the marginal velocity distribution function $\rho_{v,\omega}\left( \left\{ \bm{v}_i, \bm{\omega}_i   \right\} \right)=\int \prod_i d\bm{r}_i \rho(\Gamma)$ is well described by the product of two Gaussians for the velocities and angular velocities, respectively. } 
We make an ansatz for
the stationary $N-$particle
distribution, which neglects correlations between positions and
velocities and assumes independent particle velocities:
\begin{equation}\label{rho}
\rho(\Gamma):=  \rho\left( \left\{ \bm{v}_i, \bm{r}_i, \bm{\omega}_i   \right\} \right) =
  W_N(\{\bm{r}_i\})\prod_i \rho_{v,\omega}\left( \bm{v}_i, \bm{\omega}_i   \right)
\end{equation}
The function $W_N(\{\bm{r}_i\})$ gives zero weight to overlapping configurations and is 1 otherwise. Below we will show that velocities and positions are indeed correlated, even over long distances. But the simulation data also suggest, that the magnitude of these long ranged correlations is {smaller than the steric constraints encoded in $W_N$}. Here we assume, that it is possible to ignore these contributions to the {state variables}  we want to calculate with our theory. This approximation immediately leads to the expectation that structural quantities remain identical to the SD case at the same packing fraction; however see Sect.~\ref{sec:VI}. The velocity correlations, also neglected in the theory, are studied in Sect.~\ref{sec:VII}.

\section{State variables, equipartition and equation of state}
\label{sec:V}

The equilibrium state of a simple fluid is fully characterized by its
temperature, $T$, pressure, $p$ and packing fraction $\phi$. Furthermore equipartition of
energy holds and an equation of state relates the three state variables.
We show here that the stationary state of our system, which respects all conservation laws of a simple fluid but breaks time inversion symmetry, violates equipartition and  no unique relation connects $T,p$ and $\phi$. 

\subsection{Violation of Equipartition}
\begin{figure}[ht]
    \centering
    \includegraphics{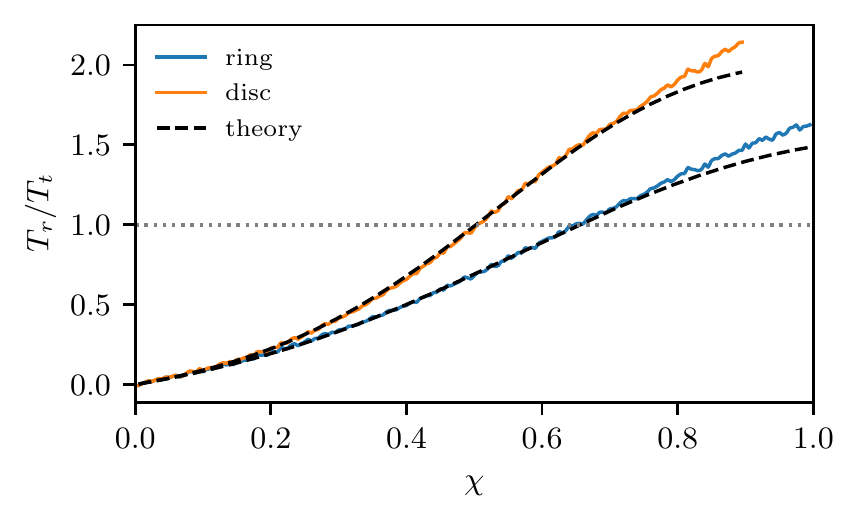}
    \caption{Ratio of rotational, $T_r$, to translational energy, $T_t$, as a function of $\chi$ for packing fraction $\phi=0.6$; results of the simulation are shown in orange (discs, $\alpha=2$) and blue (rings, $\alpha=1$) in comparison to the analytical theory (dashed line).
      The dotted line corresponds to quasi-equipartition.
    }
    \label{fig:eT_chi}
\end{figure}

Since the scattering law breaks time reversal symmetry, detailed balance is
violated. Therefore the expected probability distribution of
velocities in the stationary state will deviate from the
Maxwell-Bolzmann distribution of thermodynamic equilibrium.  To
investigate these deviations, we measure the velocity variances, i.e.~the
rotational and translational energies of the system. In the case of
thermodynamic equilibrium the equipartition theorem holds and gives
$\frac{1}{2}T_0$ as energy per degree of freedom. In contrast to that,
the energies in the time reversal symmetry violating system depend on the density as well as
on kinetic parameters,
especially the moment of inertia and $\chi$. The $\chi$-dependence is shown in
Fig.~\ref{fig:eT_chi}; it is different for rings ($\alpha=1$) and disks ($\alpha=2$). The effective temperature for the translational
degrees of freedom is denoted by $T_t= \frac{1}{N}\sum_{i=1}^Nm \langle \bm{v}_i^2/2\rangle$, the one for the rational degrees by $T_r=\frac{1}{N}\sum_{i=1}^NI \langle \bm{\omega}_i^2\rangle$.
Since energy is conserved\add{, including throughout the transients to the stationary state,} $T_r$ is
determined by $T_t$ and the initial temperature $T_0$ via
\begin{equation}
    T_r = 3T_0 - 2 T_t.  \label{eq:detTr}
\end{equation}

For values of $\chi$ close to zero,
 Fig.~\ref{fig:vn} showed that particles depart faster
after the collision as compared to the smooth case.
Rotational energy is
transferred to translational  motion
leading to a lowering of the rotational energy for small values of $\chi$.  During relaxation into the stationary state, the transfer of energy from rotation to translation slows down with reducing $\chi$ because of the smaller momentum transfer in  Eq.~\eqref{eq:deltap3}. This causes the relaxation process to become arbitrarily slow for $\chi\to0$ (not shown). Consequently,  the stationary state depends non-analytically on  $\chi$ in the limit of $\chi\to0$. The equilibrium case of smooth disks, valid for $\chi=0$, is not attained in the limit of $\chi\to0$; rather rotations {stop completely}. 
For large  $\chi$, viz.~for $\chi$ close to $1$,  sticky collisions dominate,  and 
 the postcollisional
velocities $v_n'$ are smaller than in the smooth case, see Fig.~\ref{fig:vn}. Translational energy is transferred to rotational modes, and
the translational energy decreases, while the rotational energy increases. Between these two cases, there is a value for $\chi$ with equal rotational and translational temperatures. We refer to this specific choice of $\chi$ as quasi-equipartition.

To compare the results of our simulations with the predictions of the analytical theory (Eqs.~\ref{coll_op},\ref{rho}), we compute \add{the change of the translational energy per particle}:
\begin{gather}
   \partial_tE_{t}= 0= \frac{1}{2} \sum_{i\neq j} \int \diff \Gamma \rho\lb \Gamma \rb i \mathcal{T}_{ij} \frac{1}{N} \sum_{k=1}^N \frac{m}{2} \bm{v}_k^2.
 \end{gather}

 \add{Setting this time derivative to zero determines the stationary state \cite{Huthmann1997}. Because of our ansatz for the stationary pdf and Eq.~\eqref{eq:detTr}, a single equation for the translational temperature arises, which is solved by a unique result depending on the kinetic parameters, $T_t=T_0\;\hat{T}_t(\chi,\alpha)$. The result is determined by the balance between energy transfer from rotational to translational kintetic energy and vice versa. In theory, $T_t$ is not a  function of the packing fraction, while the simulations show that it varies by less than 2\% for $\phi\le0.6$. }
 Details of the calculation of $\hat{T}_t(\chi,\alpha)$ are given in the Appendix. Here we just
 show the results for the stationary translational and rotational temperature of rings and discs in
 Fig.~\ref{fig:eT_chi}. The agreement with the simulations is good for
 both moments of inertia \add{ but worsens for $\chi$ approaching its maximal value}. Furthermore the sensitive dependence on
 $\chi$, as discussed above is well captured by the simple ansatz
 \eqref{rho}. \add{Reassuringly the theoretical calculation requires no parameter to be taken from the simulations.}\\

 \subsection{Pressure}

In the  MD simulations the pressure can be calculated  by keeping track of all collisions and their momentum transfers in a given time window. 
We start from the microscopic formulation of the symmetrized stress tensor~\cite{Hansen} 
\begin{gather}
    \sigma_{\alpha\beta} V=  \sum_i (m v_i^\alpha v_i^\beta)  + \frac{1}{2} \sum_{ij} \left[
    (f_{ij})_{\alpha} (r_{ij})_\beta  +  
    (f_{ij})_{\beta} (r_{ij})_\alpha \right]  ,  \label{eq:stressT}
\end{gather}
where the first summand, the kinetic part, will be neglected because it is trivially connected to the translational energy and can be added if needed.
For hard spheres the time averaged potential part is replaced by the mean momentum transfer in a given time interval $\Delta t$
\begin{gather}
    \bm{f}_{ij}(t) = \frac{1}{\Delta t} \sum_{\tau}\pv_{ij}^\tau, \label{eq:stressF} 
\end{gather}
where $\tau$ indexes all collisions in the time interval $\tau \in [t, t + \Delta t]$.
In the simulation $\Delta t=0.5\,t_0$ was used.

\begin{figure}[h]
    \centering
    \includegraphics{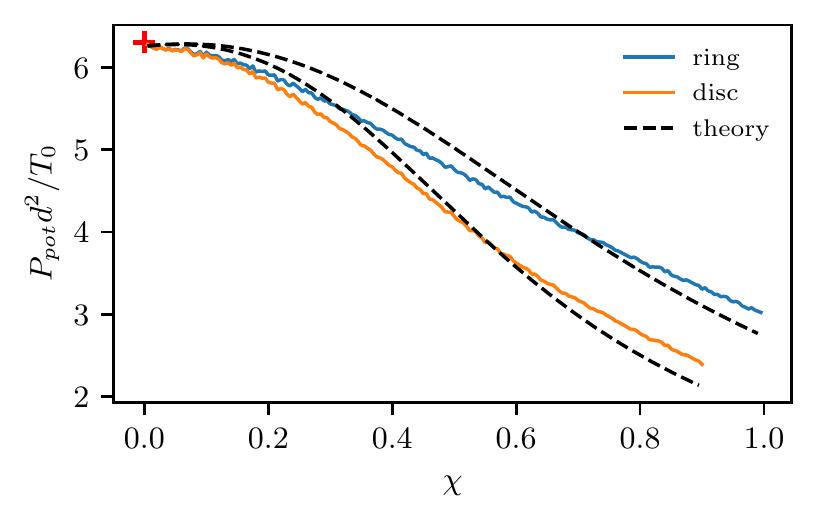}
    \caption{Pressure in the stationary state from simulations as compared to analytical theory for packing fraction $\phi=0.6$. For intermediate $\chi$, where the system is close to quasi-equipartition, the approximate theory is good. For  small and large values,  deviations occur due to underestimating correlated velocities. The red cross denotes the Baus-Colot value of smooth discs at $T=1.5\,T_0$. }
    \label{fig:eP_chi} 
\end{figure}

In Fig.~\ref{fig:eP_chi}, we display the pressure for rings and discs
as a function of $\chi$, as obtained from simulation. The pressure is
seen to decrease with $\chi$, because of two reasons. The first is that the sticky behavior of
collisions is more pronounced for larger $\chi$ and can be seen as an
effective attractive interaction. The second is the decrease in translational energy with increasing $\chi$, leading to a lowered collision frequency.

In the analytical theory, we follow Martin et al.~\cite{Martin1972} who define the momentum density in a system with rotational degrees of freedom as
\begin{gather}
    \bm{\vartheta}(\bm{q}) = \sum_{j=1}^N \left( m \bm{v}_j + i \frac{I}{2} \bm{q} \times \bm{\omega}_j  \right) e^{-i \bm{q} \cdot \bm{r}_j}, \label{eq:theta}
  \end{gather}
and compute  
 \begin{equation}
\partial_t \vartheta_{\alpha}=i \mathcal{L}_+\vartheta_{\alpha}=-iq_{\beta}\sigma_{\alpha\beta} \label{eq:conti}
\end{equation}
with help of the Pseudo-Liouville operator and the approximate $\rho(\Gamma)$. \add{This also verifies the symmetric stress tensor given in Eqs.~(\ref{eq:stressT},\ref{eq:stressF}).}

The Pseudo-Liouville-approach results in the equation for the pressure (see the Appendix)
\begin{eqnarray}
     &&P_{pot}=  2\,d^2 \, n^2 \, g(d) \, T_t \, \sqrt{\frac{ T_t}{T_t+T_r\alpha }} \int_{0}^{\frac{\pi}{2}} \text{d} \gamma      \coss{\gamma} \coss{\chi\gamma} \n \\
    && \lb  \coss{\gamma}^2+\frac{T_t}{T_t+T_r\alpha} \sinn{\gamma}^2  \rb^{-2}  \frac{\coss{\gamma- \chi\gamma}}{1+\alpha \sinn{\chi\gamma}^2}.  \label{eq:pres}
\end{eqnarray}
The particle number density is defined as $n=\frac{N}{V}$.
Besides the fixed system parameters $\alpha$ and $\chi$\add{, the value of the stationary translational temperature} $T_t$ enters the equation, which has been computed and discussed in the previous section. $T_r$ can be calculated from Eq.\,\eqref{eq:detTr}.
The rdf value at contact $g(d)$ is calculated with the theory of Baus-Colot  for smooth disks \cite{Baus1987}. The comparison between theory and simulation is included in Fig.~\ref{fig:eP_chi}.
The red cross in Fig.~\ref{fig:eP_chi} marks the Baus-Colot result for a SD liquid at a temperature of $T=1.5\,T_0$. This indicates that in the $\chi\to 0$ limit the macroscopic thermodynamic properties of the T-violating disks are similar to smooth (non-rotating) disks.

\subsection{Equation of state}
\begin{figure}
    \centering
    \includegraphics{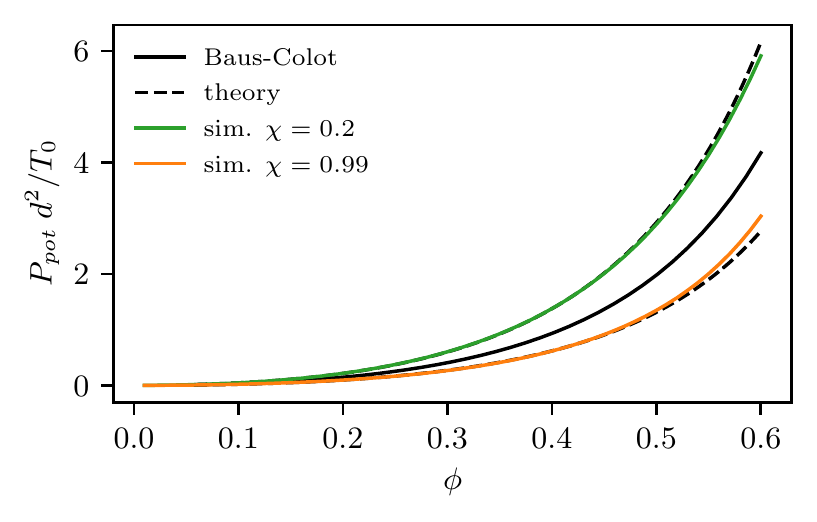}
    \caption{Equation of state of rings ($\alpha=1$) for two values of $\chi$ in comparison to the analytical theory. Also shown (black line) is the Baus-Colot theory for smooth hard disks. \add{Dashed black lines give the theory of Sect.~IV for the two $\chi$ values. }}
    \label{fig:equState}
\end{figure}

The equation of state for the T-violating system is shown in Fig.~\ref{fig:equState}. Simulation results are compared to results from the approximate analytical theory Eq.\,\eqref{eq:pres}. While in equilibrium the pressure is a function of thermodynamic parameters only, viz.~$T$ and packing fraction $\phi$ for SD, in the stationary state, kinetic parameters affect $P$, which becomes a function $P= n T_0\; \hat{P}(\phi,\chi,\alpha)$. Fig.~\ref{fig:equState} shows its dependence on $\phi$ and $\chi$, Fig.~\ref{fig:eP_chi} its dependence on $\chi$ and $\alpha$, the rescaled inverse moment of inertia. The increasing importance of sticky collisions for increasing $\phi$ and $\chi\approx 1$ explains the lowering of the pressure akin to an effective attraction. The theory captures these trends well.

For high packing fractions the difference between theory and simulation increases. That can be traced back to the fact, that all the correlations are introduced by instantaneous collisions. The rate of collision increases with increased density, so the effect of the T-violating collision rule gets bigger. 
 
In the case of quasi-equipartition at $\chi=0.654$,  the  equation of state of smooth hard disks (SD) is approached  (curve not included in Fig.~\ref{fig:equState}).

\section{Static correlations}
\label{sec:VI}

\begin{figure}
    \centering
    \includegraphics{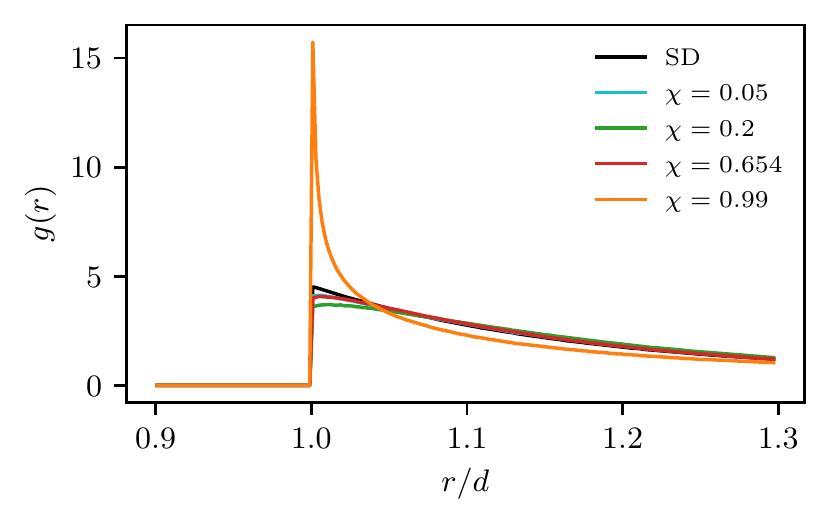}
    \caption{Radial distribution function for several values of $\chi$, where $\alpha=1$, compared to smooth disks (SD). For $\chi=0.654$ pseudo-equipartition holds (see Fig.~\ref{fig:eT_chi}). }
    \label{fig:rdf1}
\end{figure}

The scattering parameter $\chi$ affects the likelihood of particles to remain close after collisions; see Fig.~\ref{fig:vn}. This influences the local structure in a similar way as a short ranged attraction.
Sticky collisions are known to give rise to an increased probability for particles to be in contact. We thus expect a similar phenomenon here, even though $\chi$ is a  kinetic parameter. 
Fig. \ref{fig:rdf1} shows the radial distribution function \cite{Hansen} for
different values of $\chi$ at a fixed packing fraction, $\phi=0.6$. We
indeed observe a strong increase in the radial distribution at contact
as $\chi$ increases towards 1. The high contact value can be traced
back directly to the high probability of sticky collisions for
$\chi\to 1$, as depicted in Fig.~\ref{fig:vn}. For small values of $\chi$, the contact value is reduced as compared to smooth disks, because the particles move more rapidly apart after colliding - as compared to the smooth case.

\begin{figure}
    \centering
    \includegraphics{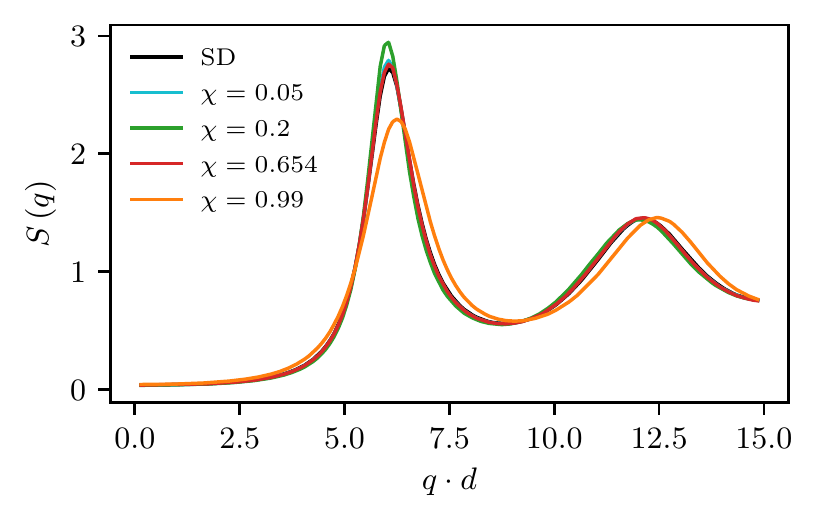}
    \caption{Static structure factor for rings ($\alpha =1$) with varying $\chi$ compared to smooth disks (SD). The $\chi=0.654$ curve hides the SD one. }
    \label{fig:staticS}
\end{figure}

In Fig.~\ref{fig:staticS} we show the
static structure factor \cite{Hansen}. The first
peak, that is connected to the average particle distance, is shifted to
higher $q$ values for the nonequilibrium system due to the sticky
behavior at large $\chi$.

\add{Figures \ref{fig:rdf1} and \ref{fig:staticS}, which are typical for other system parameters which we explored, show that the structure of the system remains that of a simple fluid even in the case of  $\chi\to0$ and $\chi\approx1$, where equipartition is violated clearly.}
 
\section{Velocity Correlations}
\label{sec:VII}

In the context of active self-propelled particles the emergence of velocity correlations in space is a known phenomenon \cite{Marconi2016,Flenner2016}. The T-violating liquid is not consuming energy, but breaks T-symmetry on a microscopic level and also shows velocity correlations in space. 

The collective velocity correlations in real space are defined as {(Note the prefactor which renders $g$ dimensionless, and finite for $n\to0$.)}
\begin{align}
   g_{\alpha \beta} \br{r} &= \frac{m}{T_0 \, N \, n}  \sum_{i\neq  j} \; \langle v_i^\alpha v_j^\beta\, \delta\br{\bo{r} - \bo{r}_{ij} } \rangle\;.
\end{align}
They arise from correlations among different particles, $i \ne j$, only.
Because of isotropy and a-chirality, velocity fluctuations can be decomposed into longitudinal and transverse  components relative to the distance vector between particle pairs, which leads to
\begin{align}
    g_{\alpha \beta} \br{r} =  g^L\br{r}\;  \hat{r}_\alpha \hat{r}_\beta  + g^T\br{r}\; \br{\delta_{\alpha \beta} - \hat{r}_\alpha \hat{r}_\beta }, \label{eq:gLgTdecomp}
\end{align}
with distance dependent isotropic functions $g^L(r)$  and $g^T(r)$.

In reciprocal space, the velocity autocorrelation of an isotropic a-chiral fluid can also be decomposed into  longitudinal ($C_L$) and transverse ($C_T$) parts\add{, which both are real functions of the wavenumber $q=|\bm{q}|$. Here the direction is set by the wavevector with $\hat{\bm{q}}=\bm{q}/q$}. The decomposition reads:
\begin{eqnarray}
  C_{\alpha\beta}(\bm{q})&=& \frac{m}{N T_0} \sum_{i,j=1}^N \mean{v_i^{\alpha}v_j^{\beta}e^{-i\bm{q}\cdot\bm{r}_{ij}}}\\
&=&                     \hat{q}_{\alpha}\hat{q}_{\beta}\; C_L(q)+(\delta_{\alpha\beta}-\hat{q}_{\alpha}\hat{q}_{\beta})\; C_T(q) \label{eq:CLCTdecomp}
  \end{eqnarray}

The connection between velocity correlations in real and reciprocal space is given by Fourier transformation:
\begin{equation}
    C_{\alpha\beta}(\bo{q}) = \frac{T_t}{T_0}\; \delta_{\alpha\beta} +  n \, \int d \bo{r} \;  e^{-i \bo{q} \cdot \bo{r}}\;  g_{\alpha\beta}(\bo{r})\;.
\end{equation}  
\add{Here, the first term arises from the single-particle contribution.}

\subsection{Local velocity correlations}

\begin{figure}
    \centering
    \includegraphics{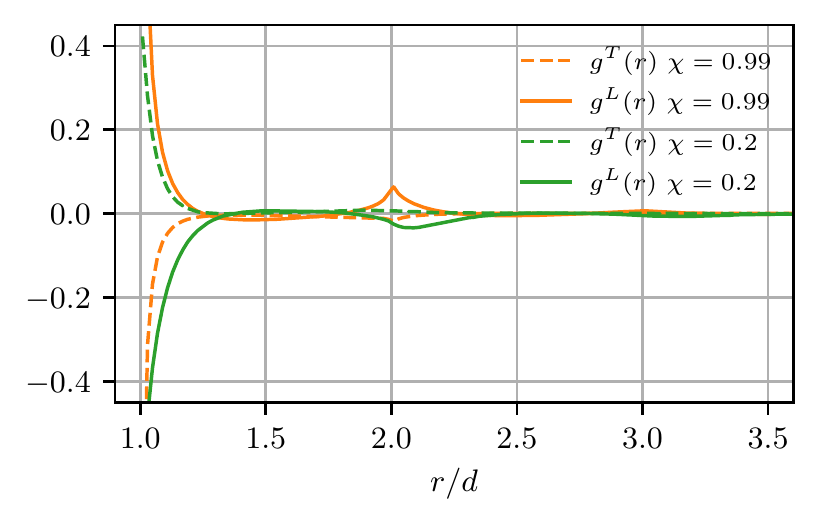}
    \caption{Velocity correlations of ring-like disks ($\alpha=1)$ in real space at packing fraction $\phi=0.6$. 
    For (smooth or rough) disks in equilibrium, the collective velocity-correlations vanish. The finite values at contact are outside of the visible range.  }
    \label{fig:vvr}
\end{figure}

 The correlation functions of longitudinal  and of transverse  collective velocity fluctuations are functions of distance.
Simulation results of $g^L(r)$ and $g^T(r)$ are shown in Fig.~\ref{fig:vvr} for two characteristic values of $\chi$. 
Frequently occurring sticky collisions at $\chi=0.99$ give rise to strong positive longitudinal correlation at $r\sim d$, while $g^T$ show less strong and negative correlations at contact.  For $\chi=0.2$ both functions change sign, compared to the corresponding functions for $\chi=0.99$. From the function values at $r\sim d$ it can be concluded, that in systems with $\chi=0.99$ particles tend to move in parallel after a collision, while for systems with $\chi=0.2$ particles tend to move away from each other quickly. The spatial dependencies indicate the fluid structure of neighbour shells surrounding individual particles \cite{Hansen}.

\begin{figure}[ht]
    \centering
    \includegraphics{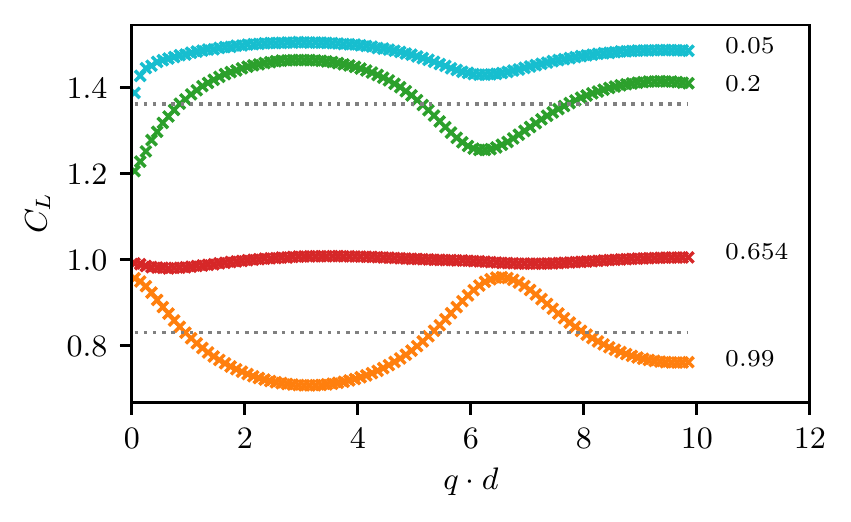}
    \caption{Longitudinal velocity correlations for rings ($\alpha=1$) at $\phi=0.6$. The right column denotes the $\chi$-values. The dashed lines give the values if no correlations were present in the system \add{for $\chi=0.2$ and 0.99}. For $q$-values around the structure peak (see Fig. \ref{fig:staticS}) the correlations are the strongest.}
    \label{fig:vvqL}
  \end{figure}

\begin{figure}[ht]
    \centering
    \includegraphics{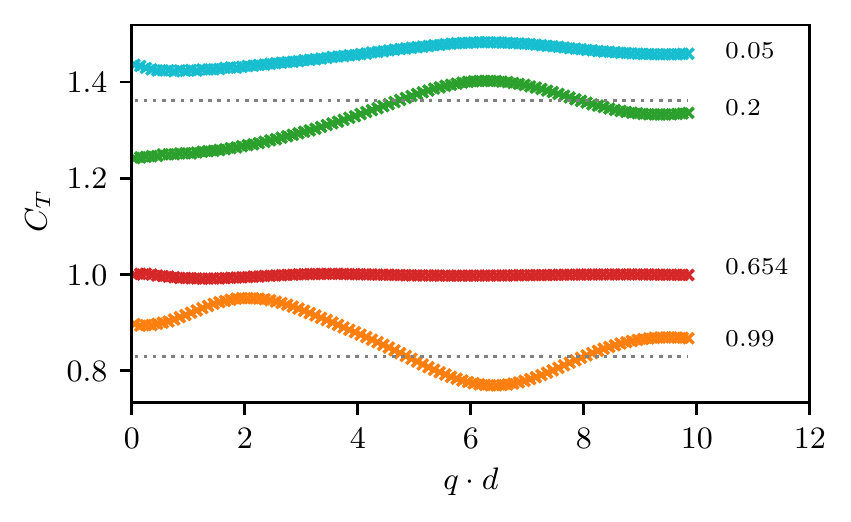}
    \caption{Transverse velocity correlations for rings at $\phi=0.6$. They show almost the inverted behavior of the longitudinal ones, but it is less strong. Again the strongest correlations occur around the peak of the static structure factor. }
    \label{fig:vvqT}
\end{figure}

Simulation results for $C_L(q)$ and $C_T(q)$ are shown in
Figs.~\ref{fig:vvqL},\ref{fig:vvqT}. Both functions are seen to oscillate as
a function of $q$ for both, small $\chi=0.2$ and large $\chi=0.99$.
The wavelength of the oscillation is comparable to the peak position of the static structure function, see Fig.~\ref{fig:staticS}. This reflects the fluid-like local structure consisting in neighbour shells; see Fig.~\ref{fig:vvr}.
If the velocities were uncorrelated, 
$C_L(q)=T_t/T_0=C_T(q)$ would hold, which is shown as dotted lines for comparison. 

  Even though we cannot compute these correlations analytically, the
  discussion of sticky collisions 
again  helps to
  get a qualitative understanding of the observed correlations.  For
  large values of $\chi=0.99$, particles in proximity tend to move
  parallel, giving rise to large positive correlations in $C_L$ around
  the peak in the structure factor.  As $\chi$ decreases, the
  oscillations weaken and disappear around the
  $\chi$-value, ($\chi=0.654$), for which quasi-equipartition was
  observed in Fig.~\ref{fig:eT_chi}. For still smaller values of
  $\chi$, e.g. $\chi=0.2$, the normal component after collision,
  $|v_n^{\prime}|$, is even larger than for the SD limit
  (Fig.\ref{fig:vn}) and the sign of the oscillations is
  reversed. Finally, as $\chi \to 0$, approaching the SD limit, the local
  correlations disappear.
The oscillations of $C_T$ are out of phase to $C_L$ with comparatively large correlations for particles in proximity and small $\chi$, when the normal component after collision, $|v_n^{\prime}|$, is even larger than for the SD limit.

\begin{figure}[ht]
    \centering
    \includegraphics{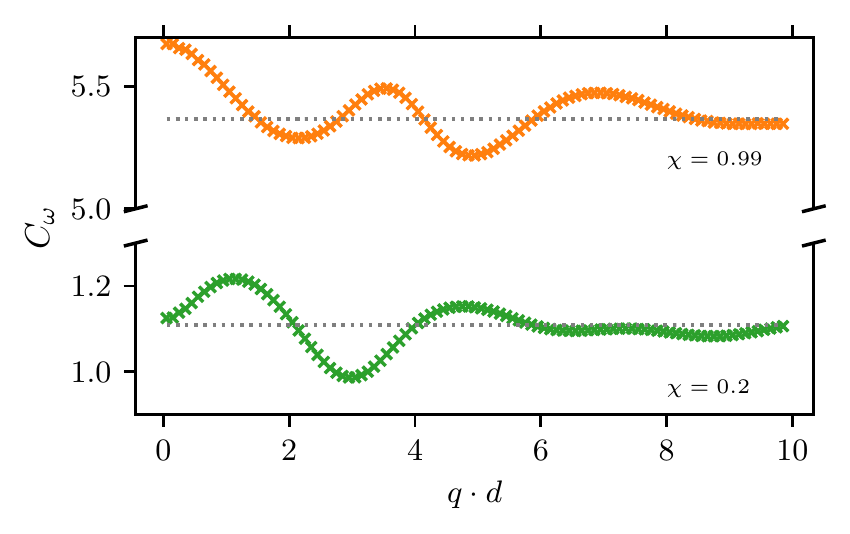}
    \caption{Correlations of angular velocities for rings at $\phi=0.6$. \add{They change sign relative to the uncorrelated value set by $T_r$ like the translational correlations with comparable magnitude, yet the characteristic length scales are much larger.}  }
    \label{fig:wwq}
\end{figure}

The magnitude of the velocity correlations in Figs.~\ref{fig:vvqL},\ref{fig:vvqT} can be compared to the magnitude of density correlations as measured by the static structure factor $S_q$ in Fig.~\ref{fig:staticS}. While the collective structural correlations $S_q-1$ are of order unity, the collective velocity correlations, $C_{L,T}-(T_t/T_0)$, are around a factor of five smaller. This may be the reason why the neglect of the correlations between velocities and positions in Eq.~\eqref{rho} adequately describes the macroscopic state variables.  

Correlations of the rotational velocity, $C_\omega(q)=\frac{m d^2}{N T_0}\sum_{i,j=1}^N \mean{\omega_i\omega_j e^{-i\bm{q}\cdot\bm{r}_{ij}}}$ also show oscillations as displayed in Fig.~\ref{fig:wwq}. The wavelength of oscillation is \add{larger than for the translational velocities. It exceeds twice} the \add{mean particle distance which characterizes the correlations in $C_L$ and $C_T$. Again,} the oscillations at high and low values of $\chi$ are approximately out of phase by 180 degrees.

The hydrodynamic momentum  field defined in Eq.~\eqref{eq:theta} also exhibits stationary correlations. The longitudinal momentum correlation function agrees with the longitudinal velocity function. The transversal momentum correlation is a linear combination of the transversal velocity correlation and the correlations of the angular velocities. The latter enters with a prefactor  $(I^2/4)\, q^2$ and thus dominates
for not too small $q$.

\subsection{Long range velocity correlations}

The stationary state shows long range correlations of the particle velocities. These correlations can be observed in reciprocal or real space, but manifest themselves differently.

The simulation data shows, see Figs.~\ref{fig:vvqL}, \ref{fig:vvqT}, that longitudinal $C_L(q)$ and transverse $C_T(q)$ velocity correlations in $q$-space approach different values for $q \to 0$.
If we denote the limits $C_T\br{ q \to 0}=C_T$ and $C_L$ analogously, the simulation data suggests the non-analytic form
\begin{gather}
    C_{\alpha\beta}\br{\bo{q} \to 0 } = \br{C_L - C_T}\frac{q_\alpha q_\beta}{q^2} + C_T  \delta_{\alpha\beta}\;. \label{eq:CLCTLim}
\end{gather}
\add{A gap, $C_L-C_T$, thus arises in the longitudinal viz.~compressional velocity fluctuations.}
The inverse Fourier transform of Eq. \eqref{eq:CLCTLim} for large distances has the form 
\begin{gather}
    \lim_{r \to \infty} \left. \int \frac{d \bo{q} }{4\pi^2} e^{i \bo{q} \cdot \bo{r}} \frac{q_\alpha q_\beta}{q^2} e^{-\epsilon q}  \; \right|_{\epsilon=0} = \frac{1}{2 \pi}\frac{1}{r^2} \br{ \delta_{\alpha \beta} - 2 \frac{r_\alpha r_\beta}{r^2}} \label{eq:qrFT} .
\end{gather}
Comparing the structure of Eq. \eqref{eq:qrFT} with Eq. \eqref{eq:gLgTdecomp}, one sees that for large distances

\begin{gather}
 \lim_{r \to \infty}   g^T(r)=-\lim_{r \to \infty} g^L(r) = \frac{1}{2 \, \pi  \, n} \, \frac{C_L - C_T}{r^2} \label{eq:powerLaw}
\end{gather}
 holds.
In real space, a power law of the form $\sim r^{-d}$ with $d=2$ is connected to different $q \to 0$ limits of $C_L(q)$ and $C_T(q)$. \add{The real space correlation $({\bf 1}-2\hat{\bo{r}}\hat{\bo{r}})/r^2$ has vanishing rotation.}

\begin{figure}
    \centering
    \includegraphics{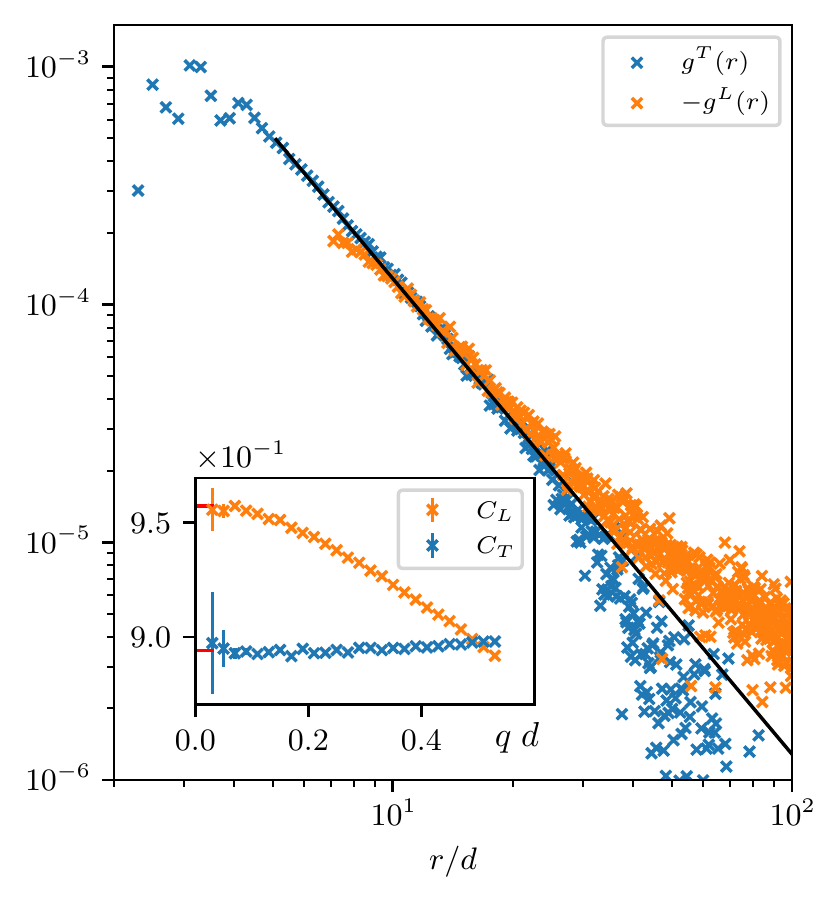}
    \caption{
    The velocity correlations $g^L(r)$ and $g^T(r)$ 
    exhibit a power law  tail in the far field (at N=250k, $\chi=0.99$, $\phi=0.6$). The black line  shows a fit of the asymptotic form $f(r)=a/r^2$ on the data of $g^T$, based on Eq.~\eqref{eq:powerLaw}. Deviations for large $r$ are due to finite size effects discussed in App.~\ref{sec:fintieSize}. In the inset, the amplitude of the power law (red bars) is compared to the gap  in the velocity correlation functions for $q\to0$.
    }
    \label{fig:vvrTN}
\end{figure}

Figure \ref{fig:vvrTN} shows the simulation results for the velocity correlations in real space in a system where $\chi=0.99$. The power-law tails have a very small magnitude in real space as can be seen in comparison to Fig. \ref{fig:vvr}, showing the much stronger velocity correlations for the first shells. 
The black curve shows a fit of the form $a/r^2$ on the simulation data of $g^T(r)$, where the parameter $a$ is connected to $C_L-C_T$ 
via Eq. \eqref{eq:powerLaw}. A fit on $g^L$ is a second independent way of determining this gap. The inset in Fig.~\ref{fig:vvrTN} focuses on the small wavevector limits of $C_L(q)$ and $C_T(q)$. 
The distance between the red bars in the inset was calculated from the mean of the two power-law amplitudes fitted in real space. It matches well the gap from the directly measured data in q-space.

For the value of $\chi=0.99$ the consistency between simulation data in real and reciprocal space is quite convincing. Smaller values of $\chi$ are harder to evaluate because of finite size effects in the simulations. Appendix \ref{sec:fintieSize} gives details on the analysis.
\begin{figure}
    \centering
    \includegraphics{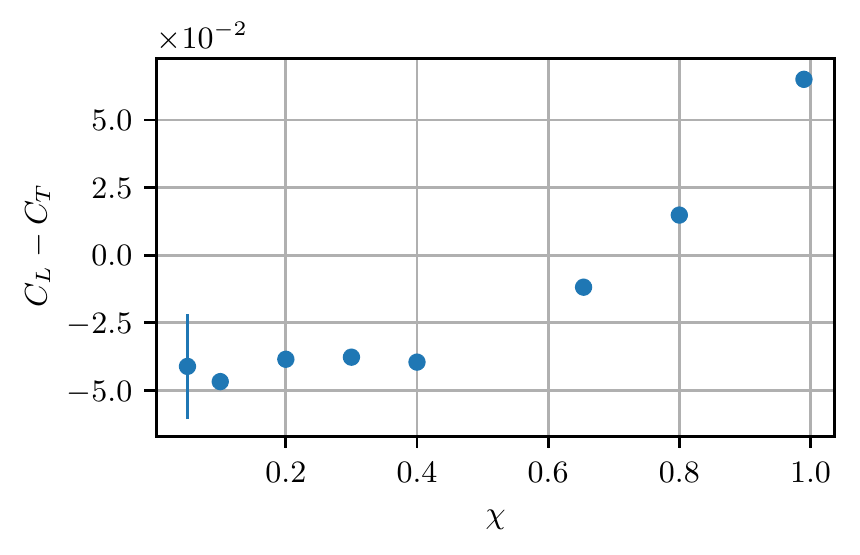}
    \caption{The amplitude of the long ranged velocity correlations. A change of sign is observable around the value of quasi-equipartition. The top and bottom of the error bars denote the coefficients from a fit to $g^T$ and $g^L$; for details of the analysis see App.~\ref{sec:fintieSize}.  The largest error bar at $\chi=0.05$ is included. }
    \label{fig:deltav}
\end{figure}
The algebraic decay of the velocity correlations characterizes the stationary states for all couplings between rotations and translations. Figure  \ref{fig:deltav} shows the amplitude $C_L-C_T$ determined consistently in real and reciprocal space as function of the parameter $\chi$. As discussed above, positive and negative amplitudes result from the tendency of particles to move in parallel or oppositely depending on $\chi$. A sign changes happens around the $\chi$ value for quasi-equipartition.
Similar to the observations of the energies in Figure \ref{fig:eT_chi} and the pressure in Figure \ref{fig:eP_chi}, the measurements in Figure \ref{fig:deltav} suggest a limit of the gap $C_L-C_T$ for $\chi \to 0$ which is unequal to $C_L-C_T=0$, which holds in the case of smooth spheres $\chi=0$. 

The magnitude of the long range velocity correlations can be compared with their local counterparts for the parameters of our simulations (inter alia packing fraction $\phi=0.60$). As function of wavevector, the collective velocity correlations $C_L(q)$ or $C_T(q)$ vary around $\pm0.2$ relative to the single particle term $T_t/T_0$ (Fig.~\ref{fig:eT_chi} shows that $T_t/T_0$ can differ from unity by $100\%$). The gap $C_L-C_T$ shown in Fig.~\ref{fig:deltav} which measures the long range contribution roughly corresponds to $\pm 0.05$, which is a quarter of the total collective correlations. Clearly, the long range part of the velocity correlations cannot be considered negligible.

\section{Conclusions and outlook}
\label{sec:VIII}

Scattering laws for hard bodies allow for the possibility to break time reversal symmetry during the collision event, while all classical symmetries are obeyed. We showed that this 
leads to a stationary nonequilibrium state whose state variables and equation of state can  semi-quantitatively be described using concepts familiar from the theory of liquids.  While structural correlations remain the ones of a simple fluid,  a coupling of positions and momenta exists in the stationary state. Collective velocity correlations emerge as familiar in more complex systems like self-propelled particles or granular systems.
\add{Intriguingly, the collective velocity correlations become long ranged and asymptotically decay like a power law, $\propto 1/r^2$ in 2D.} 
\add{As a further consequence, the coupling of positions and momenta invalidates the classical equipartition theorem.}

The introduced system offers many possibilities for future studies.
Its transport processes and dynamical correlations would be of interest \cite{Epstein2020}.
Kinetic theory  could calculate the low-$\phi$ properties quantitatively \cite{Noije1998,Meanwell2017}. The long ranged velocity correlations are of interest in more complex  nonequilibrium fluids \cite{Machta1980,Dorfman1994,Szamel2021} and could be discussed in more detail in the present model. Phase transitions will lead to ordered states and, possibly, to a nonequilibrium critical point for purely repulsive interactions.
The latter may be speculated on the basis of the sticky collisions causing an effective attraction.

As pointed out by Avron~\cite{Avron}, a 2-dimensional isotropic system
with broken time-reversal symmetry can display a so called odd
viscosity \cite{Banerjee2017,Epstein2020}.
The 2-dimensional system under consideration breaks time reversal
symmetry and is isotropic but achiral. A possible generalization of the scattering law can be explored breaking additional symmetries, e.g.~parity, in a
controlled fashion.   This would lead to an odd fluid, viz.~a chiral
fluid breaking time-reversal symmetry, yet conserving energy.

\begin{acknowledgements}
MF and NG thank A. Laganapan for support in the initial stages of the project and NG acknowledges funding by the DFG within SFB 1432 in project C07. 
\end{acknowledgements}

\appendix
\section{Details of the analytical theory}
\label{sec:A}

\subsection{Stationary state temperatures}
Following \cite{Aspelmeier2001}, the rate of change of the translational energy reads \add{based on the ansatz \eqref{rho}}
\begin{align}
    \partial_t E_t =& g(d)
    \frac{N}{V^2} 
    \left( \frac{m}{2 \pi T_t} \right)^2 \frac{I}{2 \pi T_{r}} \int \text{d} \omega_1 \text{d} \omega_2 \text{d} \bm{r}_1 \text{d} \bm{r}_2 \text{d} \bm{v}_1 \text{d} \bm{v}_2 \nonumber  \\
    & \exp \left( -\frac{m}{2 T_t} \left( \bm{v}_1^2 +\bm{v}_2^2 \right) - \frac{I}{2 T_{r}} \left( \omega_1^2 + \omega_2^2 \right)   \right) \nonumber \\
    &|\bm{v}_{12} \cdot \nv|\;\Theta(\bm{v}_{12}\cdot \nv)\; \delta(|\bm{r}_{12}|-d)\;\Delta E_t . \label{eq:en1}
\end{align}

The crucial assumption of uncorrelated velocities enters as the choice of independent Gaussian distributions of the velocities, but with two temperatures $T_t$ and $T_r$ for translational and rotational degrees of freedom. $\Delta E_t$ describes the difference in translational energy before and after the collision.  Transformation into relative coordinates  
\begin{gather}
    \omega = \frac{\omega_1+ \omega_2}{\sqrt{2}}, \qquad \Omega = \frac{\omega_1-\omega_2}{\sqrt{2}} , \nonumber \\
     \bm{v} = \frac{\bm{v}_1- \bm{v}_2}{\sqrt{2}}, \qquad \bm{V} = \frac{\bm{v}_1+\bm{v}_2}{\sqrt{2}} , \nonumber \\
      \bm{r} = \bm{r}_1- \bm{r}_2, \qquad \bm{r}_1 = \bm{r}_1 ,  \label{eq:subs}
\end{gather}
simplifies the problem. The resulting integral from Eq.\,\eqref{eq:en1} can be solved trivially for $\bo{r}_1$ (box volume), $\bo{\Omega}$ and $\bo{V}$ (Gaussians). The integral over $\bo{r}$ cancels the Dirac-$\delta$.
In these coordinates $\Delta E_t$ reads
\begin{align}
    \Delta E_t =&  \frac{m}{2} \left( \bm{v}_1'^2 + \bm{v}_2'^2 - \bm{v}_1^2 - \bm{v}_2^2 \right)  \n \\
    =& \frac{1}{m}\Delta \bm{p}^2 + \sqrt{2} \bm{v} \cdot \Delta \bm{p}.
\end{align}

The momentum transfer $\Delta \bo{p}$ is given by the collision rule Eq.\,\eqref{eq:deltap1}. 

In polar coordinates as shown in Figure \ref{fig:colEvent}, where we chose the $x$-axis in direction of $-\nv$, $\Delta E_t$ reads
\begin{align}
\Delta E_t =&   m \frac{ \cos(\gamma-\gamma \chi  )}{ 1+\alpha \sin(\chi \gamma)^2 } \times \n  \\ &\left[  \frac{ \cos(\gamma-\gamma \chi  ) v_c^2}{ 1+\alpha \sin(\chi \gamma)^2   }  - \sqrt{2}  v_c \lb  \cos\lb \gamma \chi  \rb v_x + \sin \lb \gamma \chi   \rb v_y    \rb  \right] .
\end{align}
The variable $\omega$ is substituted by $z=v_y - R \omega$. 
In this coordinate frame
\begin{gather}
v_x = \frac{v_c \coss{\gamma}}{\sqrt{2}}, \qquad z = \frac{v_c \sinn{\gamma}}{\sqrt{2}} , \label{eq:polar}
\end{gather}
hold. 
What is left are the integrals for $v_c$, $v_y$ and $\gamma$, where the former two can be solved as Gaussian integrals. 

With that the integral of Eq.\,\eqref{eq:en1} finally results in

\begin{widetext}
\begin{eqnarray}
    \partial_t E_t =&& - \kappa \sqrt{\frac{1}{T_t(T_t+T_{r}\alpha) }} ~ 
   \int ^{\frac{\pi}{2} }_{0}  \diff \gamma     ~    \cos(\gamma) 
       \lb \frac{ \sin(\gamma)^2 }{T_r \alpha} \lb 1-  \frac{T_t}{T_t+T_{r}\alpha } \rb  +\frac{\cos(\gamma) ^2}{T_t}     \rb ^{-\frac{5}{2}} \n  \\
   &&  \frac{ \cos(\gamma-\gamma \chi  ) }{ 1+\alpha \sin(\chi \gamma)^2 } \left[  \frac{ \cos(\gamma-\gamma \chi  ) }{ 1+\alpha \sin(\chi \gamma)^2   }  -  \lb  \cos\lb \gamma \chi  \rb  \cos\lb \gamma \rb  + \sin \lb \gamma \chi   \rb  \sin(\gamma)  \frac{ T_t }{T_t+T_{r}\alpha }  \rb  \right]. \label{eq:Tfin}
\end{eqnarray}
\end{widetext}
The variable $\kappa$ gathers constants that are of no concern, because the stationary state energies are given by the root of Eq.~\eqref{eq:Tfin}. \add{Inspection of the integrand gives the dependence of $T_t$.}

\subsection{Pressure}
The time evolution of the momentum density defined in Eq.~\eqref{eq:theta} due to the collision part in Eq.~\eqref{coll_op} reads \cite{Aspelmeier2001}
\begin{widetext}
\begin{eqnarray}
\mean{ i \mathcal{L}_+' \vartheta_\alpha } =& \frac12 \mean{  \sum_j e^{-i \bo{q} \cdot \bo{r}_j }   \sum_{k \neq l}  \mathcal{T}^{(kl)}_+ \left( m v_j^\alpha+ i \frac{I}{2} (\bo{q} \times \bo{\omega}_j)^\alpha \right) }  \n \\
=& \frac12 \sum_{k \neq l} \int \text{d} \Gamma~ \rho(\Gamma) \; (-\hat{\bm{r}}_{kl} \cdot \bm{v}_{kl}  ) \;\Theta (- \hat {\bm{r}}_{kl} \cdot \bm{v}_{kl} )\;\delta (r_{kl} - d )  \n \\
& \left[   \Delta p_\alpha \left( e^{-i \bm{q} \cdot \bm{r}_k}  - e^{-i \bm{q} \cdot \bm{r}_l}      \right) -  \left( e^{-i \bm{q} \cdot \bm{r}_k}  + e^{-i \bm{q} \cdot \bm{r}_l}      \right)  \frac{i}{2} R \left(  \hat{\bm{r}}_{kl} (\bm{q}  \cdot \Delta \bm{p})  - \Delta \bm{p}(\bm{q}\cdot \hat{\bm{r}}_{kl}) \right)_\alpha  \right]  . 
\end{eqnarray}
\end{widetext}

Expanding this equation up to linear order in $q$ around 0 gives a factor $-i\Delta p_\alpha \bm{q} \cdot \bm{r}_{kl}$ and the second bracket gives a factor 2. 
From the continuity equation \eqref{eq:conti} we obtain an expression for the stress tensor 
\begin{align}
    \sigma_{\alpha \beta} =& -\frac{n^2\, d}{4}  \int  \text{d} \bm{v}_1 \text{d} \bm{v}_2 \text{d} \bm{r}_1 \text{d} \bm{r}_2  \text{d} \omega_1 \text{d} \omega_2 \n \\
    & g(r) \delta(r - d) \rho_{v,\omega}(\bo{v}_1,\bo{v}_2, \omega_1,\omega_2)(\hat{\bo{r}} \cdot \bo{v}_{12}) \n  \\
    &\Theta(-\hat{\bo{r}} \cdot \bo{v}_{12}) \left[ \Delta p_\alpha    \hat{r}_\beta  + \Delta p_{\beta}    \hat{r}_{\alpha}  \right]  . \label{eq:sig2}
\end{align}
The diagonal entries give the normal pressure, where $xx$- and $yy$-contributions are equivalent. The ansatz for the probability distribution $\rho_{v,\omega}$ is again a product of independent Gaussians.
With the substitutions Eq.~\eqref{eq:subs}, the Eq.~\eqref{eq:sig2} becomes 
\begin{align}
 2P_{pot} =& -\frac{d\;n^2}{4}\,\int \text{d} \bo{r} \text{d} \bo{v} \text{d} \omega (  \sqrt{2} \bo{v} \cdot \hat{\bo{r}})\,g(r)\,\Theta(- \bo{v} \cdot \hat{\bo{r}} ) \delta(r-d)  \n \\
& \lb \frac{m}{2 \pi T_t} \rb  \lb \frac{I}{2 \pi T_{r}} \rb^{\frac{1}{2}}  \exp\left[-\frac{m\bm{v}^2}{2T_t} -\frac{I\omega^2}{2T_{r} }  \right] 2 \pv \cdot \hat{\bo{r}} . 
\end{align}
Again collisions along the $x$-axis are considered, and $z$ is introduced as above.
With an integration over $\bm{r}$ it follows
\begin{align}
 &P_{pot} =\frac{d n^2g(d)}{2} \int_{-\infty}^{\infty}  \text{d} v_y \int_{-\infty}^{0}  \text{d} v_x \int_{-\infty}^{\infty} \text{d} z ~  v_x   \lb \frac{m}{ T_t} \rb  \lb \frac{I}{ \pi T_{r}} \rb^{\frac{1}{2}}  \n \\
& \exp\left[-\frac{m v_x^2}{2T_t} - \frac{m z^2}{2T_{r}\alpha } -  v_y^2  \left( \frac{m}{2T_t} + \frac{m}{2T_{r}\alpha}  \right)+v_y\frac{m\,z }{T_{r} \alpha} \right] \Delta p_x .  \label{eq:P3}
\end{align}
The $v_y$ integral can be solved straightforwardly. 
Again choosing the polar coordinates defined in Eq.~\eqref{eq:polar}, plugging in the explicit form of $\Delta p_x$ and integrating over $v_c$, yields Eq.~\eqref{eq:pres}.

\section{Simulation details} \label{sec:AppSim}

\com{Ich würde die Diskussion des Gesamtdrehimpuls als Appendix B.2 mit einer Figur addieren. }

Except for the results in Fig.~\ref{fig:equState}, we consider  a fluid state at packing fraction $\phi=0.6$.
 For the $\chi$-dependent figures~\ref{fig:eT_chi}, \ref{fig:eP_chi} we simulated  one system with $N=19600$ particles for each value of $\chi$ and for a timespan of $2000\,t_0$ and measured the energy every $0.2\,t_0$, after equilibrating for $t=40\,t_0$.
   Here one has to be careful, because the equilibration time of the energies increases for decreasing $\chi$. For the equation of state in Fig.\,\ref{fig:equState} we simulated one system for each packing fraction with $N=4900$ for $4000\,t_0$. The structural quantities shown in Fig.\,\ref{fig:rdf1},\,\ref{fig:staticS} are averages over 200 independent systems with $N=90\,000$, after equilibration of $t=300\,t_0$.\\
To generate the $q$ data points in Figs.\,\ref{fig:vvqL}, \ref{fig:vvqT} and \ref{fig:wwq} we used more than 400 independent systems of $N=250\,000$ particles, the same holds for the real space correlations in figure \ref{fig:vvr}. To improve statistics we also averaged over time. To generate the figures \ref{fig:vvrTN} and \ref{fig:deltav} we used also the 400 systems of $N=250\,000$ particles, but for these figures even more time averaging was needed.

\subsection{Equilibration of slow hydrodynamic modes}

\begin{figure}
    \centering
    \includegraphics{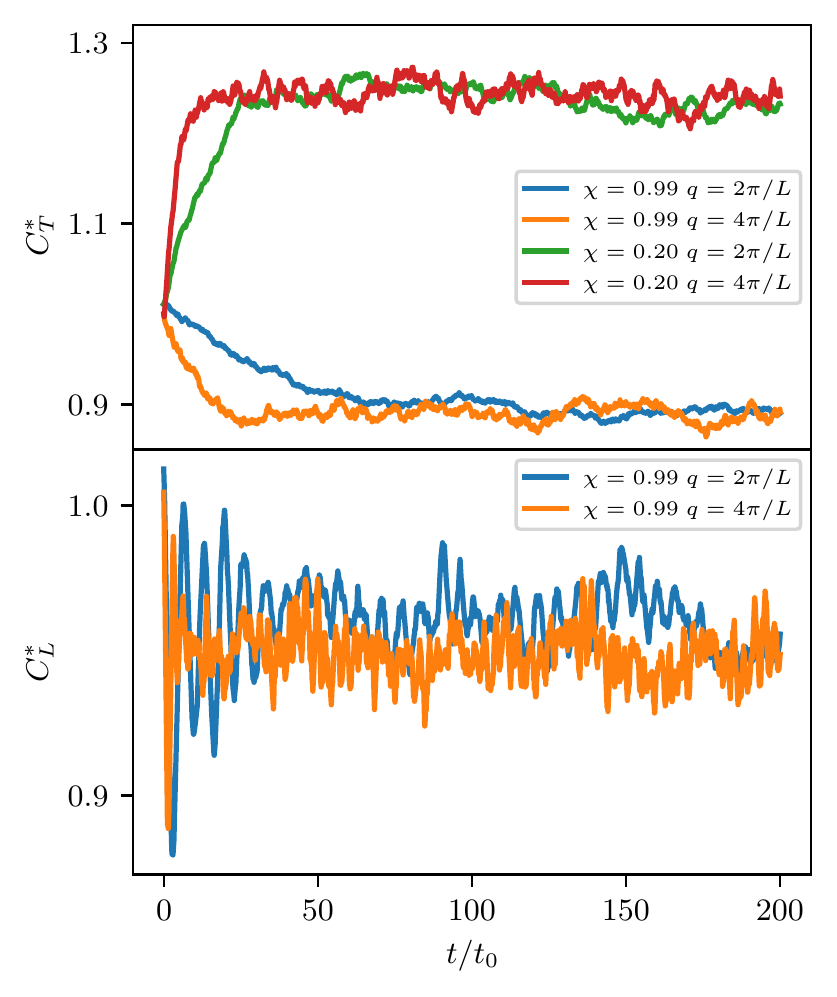}
    \caption{Transient equal time correlation of the transverse and longitudinal velocity field where $\alpha=1$, $\phi=0.6$ and $N=4900$. At $t=0$ the particles start from an amorphous structure and energies are drawn from the equilibrium pdf. Transverse correlations build up in a diffusive way, while longitudinal correlations show sound waves.  }
    \label{fig:cL0}
\end{figure}

The equilibration process of the long-wavelength velocity correlations turns out to be non trivial.  Figure \ref{fig:cL0}  shows equal time velocity correlations calculated at different timesteps during the equilibration process. At $t=0$ the structure of the system is in an amorphous state, equilibrated with the smooth disc collision rule. To reach the amorphous structure we used an inflation protocol \cite{Lubachevsky1990}. The dynamics is initialized with velocities drawn from an equilibrium Maxwell-Boltzmann distribution at $t=0$ and the collision rule is switched to the T-violating rule.
The lower panel of Fig.~\ref{fig:cL0} shows longitudinal velocity correlations.
For small wave vectors the velocity correlations show clear oscillations with a $q$-dependent frequency. This resembles sound-waves traveling trough the system repeatedly due to the periodic boundary conditions. The damping decreases with decreasing $q$, which makes equilibrating harder for small $q$ correlations.
Transverse correlations build up diffusively in the equilibration process. The upper panel in Fig.~\ref{fig:cL0} shows the transient equal time correlations for transverse velocity fields. Here again, the equilibration time increases drastically with decreasing $q$. For $q=0$ this will  not happen, since the center of mass motion is set to zero explicitly at the beginning of the simulation. \\
Figure \ref{fig:cL0} shows an averages over 11000 systems with $N=4900$\add{, which were necessary to sample the slow noise fluctuations in the hydrodynamic regime}.

\subsection{Finite size effects of the velocity correlations} \label{sec:fintieSize}
\begin{figure}[h]
    \centering
    \includegraphics{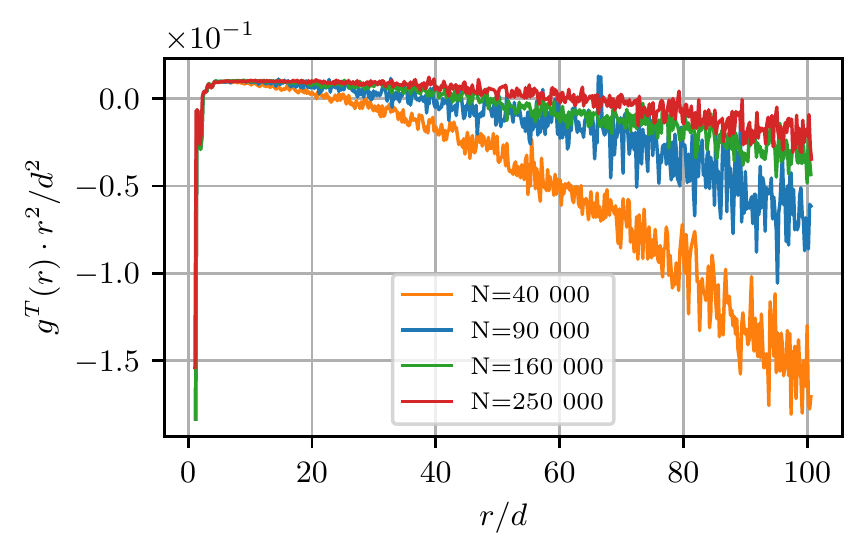}
    \caption{Finite size effects of long ranged velocity correlations at $\chi=0.99$. For increasing system sizes, the range where simulation data shows a power-law behavior increases. }
    \label{fig:vvrTNxr2}
\end{figure}

Since the decay of the velocity correlation is slow and follows a power-law, the finite size of the simulation system affects this quantity. Figure \ref{fig:vvrTNxr2} shows the slope of $r^2\, g^T(r)$ for $\chi=0.99$ and different system sizes, all at packing fraction $\phi=0.6$; at $N=40\,000$, the box size is roughly $L/d\approx228$. From this figure the agreement of the simulation data with a $r^{-2}$ power law can be read off. The larger the system is, the longer the data follows a horizontal line i.e. showing the expected $r^{-2}$ behavior. For long distances, $r^2g^T(r)$ follows a parabola, indicating that the correlations $g^T(r)$ do not decay to zero within the limits of the box size, but approach a constant value. The magnitude of this plateau value decreases with increasing system sizes. 

\begin{figure}
    \centering
    \includegraphics{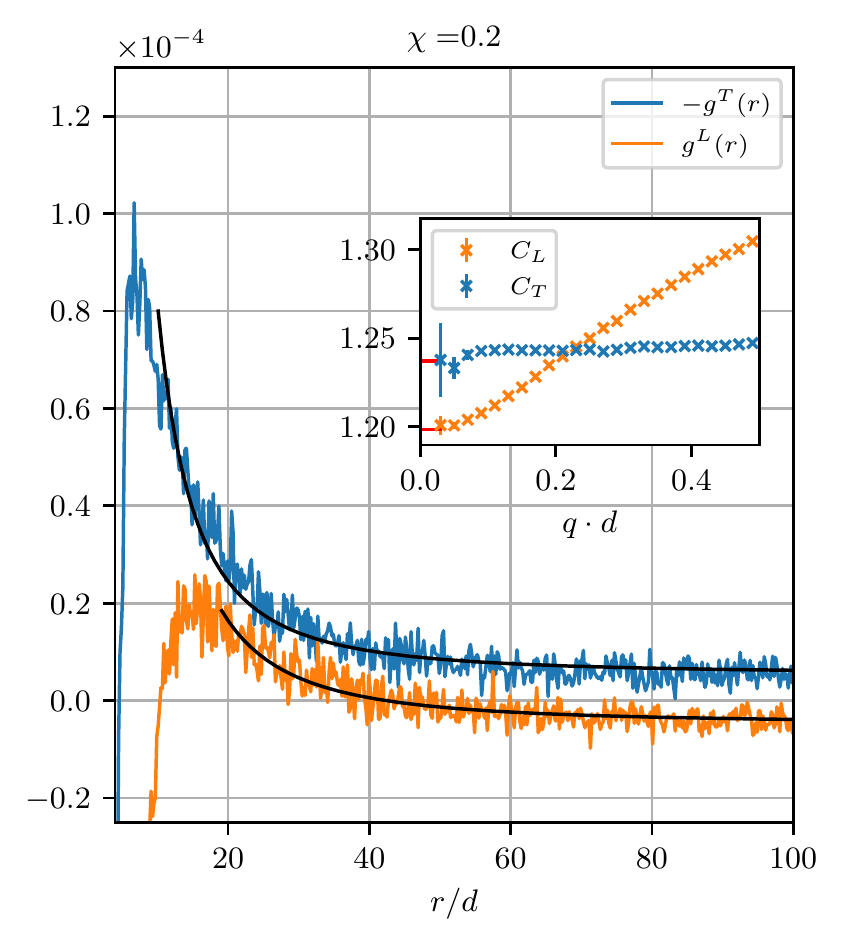}
      \caption{For small values of $\chi$, here $\chi=0.2$, the magnitude of the finite size effects can no longer be ignored.  Fits are performed with $a/r^2+c$ at $N=250\,000$. The inset compares the averaged amplitude $a$ (distance of the red lines) to the q-space data. }
    \label{fig:X20_overlap}
\end{figure}

For decreasing $\chi$ the amplitude of the power law gets smaller and the plateau arising from the finite size is not negligible anymore, see Fig. \ref{fig:X20_overlap}  for systems with $\chi=0.2$.
Moreover, the region where the power law is observed starts only from larger distances. Therefore a model of $a/r^2+c$ is used, where $c$ should account for the distortion of the power law due to finite size effects. The fit to the two curves $g^T$ and $g^L$ gives the two fit parameters $a_T$ and $a_L$. The mean of $a_T$ and $a_L$ is multiplied by $2 \, \pi \, n $  to generate a real space measurement of $C_L-C_T$. 
In this way it was possible to extract the coefficients for smaller $\chi$, shown in Figure \ref{fig:deltav}.  The error bars there show the difference between $2\pi n a_L$ and $2\pi n a_T$.

\clearpage

\bibliography{apssamp}

\end{document}